%% file: main.tex
\documentclass[sigconf, nonacm]{acmart} 

\AtBeginDocument{%
  }

\setcopyright{acmcopyright}
\copyrightyear{2023}
\acmYear{2023}
\acmDOI{XXXXXXX.XXXXXXX}
\acmConference[ASHES '23]{Manuscript submitted to the 7th Workshop on Attacks and Solutions in Hardware Security}{November 30, 2023}{Copenhagen, DK}
\acmBooktitle{Manuscript submitted to the 7th Workshop on Attacks and Solutions in Hardware Security, November 30, 2023, Copenhagen, DK}

\usepackage{placeins}
\usepackage{todonotes}
\presetkeys
{todonotes}
{
    size=\footnotesize,
    caption={},
}{}
\usepackage{siunitx}

\usepackage{subfig}

\usepackage{tikz}
\input{fig_scales}
\newcommand*\rectangled[1]{\tikz[baseline=(char.base)]{
            \node[shape=rectangle,draw,inner sep=2pt] (char) {#1};}}

\usepackage{booktabs}
\usepackage{import}

\input{glossary}

\begin{document}
\title{Modulation to the Rescue: Identifying Sub-Circuitry in the Transistor Morass for Targeted Analysis}


\author{Xhani Marvin Saß}
\affiliation{%
  \institution{Security in Telecommunications}
  \city{Berlin}
  \country{Germany}
}

\author{Thilo Krachenfels}
\affiliation{%
  \institution{Security in Telecommunications}
  \city{Berlin}
  \country{Germany}
}

\author{Frederik Dermot Pustelnik}
\affiliation{%
  \institution{Security in Telecommunications}
  \city{Berlin}
  \country{Germany}
}

\author{Jean-Pierre Seifert}
\affiliation{%
  \institution{Security in Telecommunications}
  \city{Berlin}
  \country{Germany}
}

\author{Christian Große}
\affiliation{%
  \institution{Fraunhofer IMWS}
  \city{Halle}
  \country{Germany}
}

\author{Frank Altmann}
\affiliation{%
  \institution{Fraunhofer IMWS}
  \city{Halle}
  \country{Germany}
}

\input{00-abstract/abstract.tex}
\input{01-ccsxml/ccsxml.tex}
\maketitle
\input{02-introduction/introduction.tex}
\input{03-background/background.tex}
\input{04-approach-and-setup/approach-and-setup}
\input{05-results/results}
\input{06-discussion/discussion}
\input{07-conclusion/conclusion}

\FloatBarrier
\bibliographystyle{ACM-Reference-Format}
\balance
\bibliography{./bibliography.bib}

\newpage
\input{08-appendix/appendix}

\end{document}

%% file: glossary.tex
\usepackage[abbreviations, hyperfirst=false]{glossaries-extra}
\makeglossaries

\glsdisablehyper

\newabbreviation{fi}{FI}{fault injection}
\newabbreviation{ic}{IC}{integrated circuit}
\newabbreviation{asic}{ASIC}{application-specific integrated circuit}
\newabbreviation{trc}{TRC}{time replica circuit}
\newabbreviation{tdc}{TDC}{time-to-digital converter}
\newabbreviation{mfi}{MFI}{multiple fault injection}
\newabbreviation{pch}{PCH}{Platform Controller Hub}
\newabbreviation{csme}{CSME}{Converged Security and Management Engine}
\newabbreviation{fpga}{FPGA}{field-programmable gate array}
\newabbreviation{dut}{DuT}{device under test}
\newabbreviation{cfi}{CFI}{clock fault injection}
\newabbreviation{vfi}{VFI}{voltage fault injection}
\newabbreviation{em}{EM}{electromagnetic}
\newabbreviation{emfi}{EMFI}{electromagnetic FI}
\newabbreviation{lfi}{LFI}{laser fault injection}
\newabbreviation{bbi}{BBI}{body-biasing injection}
\newabbreviation{isa}{ISA}{instruction set architecture}
\newabbreviation{risc}{RISC}{reduced instruction set computer}
\newabbreviation{por}{PoR}{power-on-reset}
\newabbreviation{slb}{SLB}{software level-based}
\newabbreviation{hlb}{HLB}{hardware level-based}
\newabbreviation{pem}{PEM}{photon emission microscopy}
\newabbreviation{eofm}{EOFM}{electro-optical frequency mapping}
\newabbreviation{eop}{EOP}{electro-optical probing}
\newabbreviation{lit}{LIT}{lock-in thermography}
\newabbreviation{fa}{FA}{failure analysis}
\newabbreviation{ir}{IR}{infrared}
\newabbreviation{nir}{NIR}{near-infrared}
\newabbreviation{sem}{SEM}{scanning electron microscope}
\newabbreviation{llsi}{LLSI}{laser logic state imaging}
\newabbreviation{soc}{SoC}{system-on-chip}
\newabbreviation{lsm}{LSM}{laser scanning microscope}
\newabbreviation{pdn}{PDN}{power delivery network}
\newabbreviation{pcb}{PCB}{printed circuit board}
\newabbreviation{sca}{SCA}{side-channel analysis}
\newabbreviation{puf}{PUF}{physical unclonable function}
\newabbreviation{emsc}{EMSC}{electromagnetic side channel}
\newabbreviation{cpu}{CPU}{central processing unit}
\newabbreviation{ich}{ICH}{I/O Controller Hub}
\newabbreviation{na}{NA}{numerical aperture}
\newabbreviation{mcu}{MCU}{micro-controller unit}
\newabbreviation{ff}{FF}{flip-flop}
\newabbreviation{psu}{PSU}{power supply unit}

%% file: 00-abstract/abstract.tex
\begin{abstract}
    Physical attacks form one of the most severe threats against secure computing platforms.
    Their criticality arises from their corresponding threat model:
    By, e.g., passively measuring an \gls{ic}'s environment during a security-related operation, internal secrets may be disclosed.
    Furthermore, by actively disturbing the physical runtime environment of an \gls{ic}, an adversary can cause a specific, exploitable misbehavior.
    The set of physical attacks consists of techniques that apply either globally or locally.
    When compared to global techniques, local techniques exhibit a much higher precision, hence having the potential to be used in advanced attack scenarios.
    However, using physical techniques with additional spatial dependency expands the parameter search space exponentially.

    \noindent
	In this work, we present and compare two techniques, namely \gls{llsi} and \gls{lit}, that can be used to discover sub-circuitry of an entirely unknown \gls{ic} based on optical and thermal principles.
    We show that the time required to identify specific regions can be drastically reduced, thus lowering the complexity of physical attacks requiring positional information.
    Our case study on an Intel H610 Platform Controller Hub showcases that, depending on the targeted voltage rail, our technique reduces the search space by around \SI{90}{\percent} to \SI{98}{\percent}. 
\end{abstract}

%% file: 01-ccsxml/ccsxml.tex
\ccsdesc[500]{Security and privacy~Hardware reverse engineering}
\ccsdesc[500]{Hardware~Transistors}


\keywords{Hardware Security, Reverse Engineering, Integrated Circuits, ASIC}


%% file: 02-introduction/introduction.tex
\section{Introduction}\glsresetall
    Physical attacks, such as \gls{fi} attacks or \gls{sca} attacks, form one of the most severe threats against secure computing platforms. Their criticality lies within their corresponding threat model.
    By, e.g., passively measuring a processing unit's environment during a security-critical operation, sensitive information may be leaked unintentionally, which leads to the disclosure of internal secrets~\cite{side-channel-power-analysis-of-an-aes-256-bootloader}.
    Moreover, by actively disturbing the physical runtime environment of an \gls{ic}, an adversary can deliberately cause specific misbehavior, which can be exploited afterward~\cite{oops-i-glitched-it-again}.
    Hence, these powerful attacks can introduce vulnerabilities into systems where from a functional point of view, none exist.

    \begin{figure}[t!]
        \centering
        \includegraphics[width=.75\linewidth]{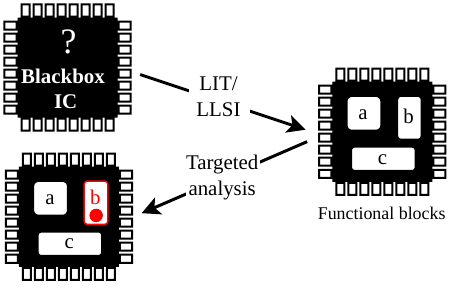}
        \captionsetup{belowskip=-12pt, aboveskip=2pt}
        \caption{Sketch of the attack flow: modulation-based techniques help identify functional blocks that other techniques can then target.}
        \label{fig:attack-flow}
    \end{figure}

    \noindent
    A way to classify different physical attack scenarios is by their area of consideration.
    While global techniques (e.g., power analysis, voltage \gls{fi}, or clock \gls{fi}) always consider the entire \gls{dut}, local techniques affect or measure only a spatially restricted area.
    Local techniques, such as \gls{lfi}, \gls{emfi}, \gls{bbi} or \gls{emsc} analysis, make highly targeted attacks viable, as the considered region can be restricted to a specific target region.
    By exclusively covering a sub-region of interest of the \gls{ic}, any side effects from and to surrounding components are avoided.
    Nonetheless, even for the class of localized techniques, there are differences to be considered: while \gls{emfi}, \gls{emsc}, and \gls{bbi} all provide means of spatial resolution, \gls{lfi} can be executed at a much finer granularity.
    In other words, the higher the spatial restriction is, the more precise an attack becomes.
    However, the gain in precision comes for the sake of complexity.
    As every additional parameter that has to be discovered during a physical attack increases the search space by exponential means~\cite{fill-your-boots,AMD_EM_fault}, introducing locality in addition to the fault's parameters (i.e., X-, Y-, and Z-position) results in a combinatorial explosion.
    If an adversary has to identify the target circuitry's position on the entire silicon die, the resulting expansion of the search space leads to the impracticability of an otherwise feasible attack.
    
    \noindent
    Due to the aforementioned expansion, a vast amount of research has been proposed to counter the expansion of search space under specific circumstances.
    \Citeauthor{breaking-black-box-crypto-devices-using-laser-fault-injection}~\cite{breaking-black-box-crypto-devices-using-laser-fault-injection} proposed a method based on optical inspection to exclude specific regions of interest.
    In addition, they proposed the exploitation of side-channel information to further narrow the search space.
    By measuring the current while stimulating the \gls{ic}'s backside with a laser, \Citeauthor{On-the-complexity-reduction-of-laser-fault-injection-campaigns-using-OBIC-measurements}~\cite{On-the-complexity-reduction-of-laser-fault-injection-campaigns-using-OBIC-measurements} showed for a given \gls{mcu}, that \glspl{ff} can be identified, which represent lucrative targets for \gls{lfi} in general.
    While previous work successfully identified areas of interest in specific circumstances, the general identification of regions on a black box silicon die still poses a hard task~\cite{AMD_EM_fault}.
    
	\noindent
    In this work, we propose the identification of sub-circuitry based on the modulation of specific, physically isolated voltage supplies.
    The modulation of a particular circuitry of interest via its voltage supply causes local physical effects, which can be measured by techniques commonly encountered in the \gls{ic} \gls{fa} domain.
    By modulating a single voltage rail while leaving the others unmodified, the external modulation manifests, e.g., in local temperature variation or a change in amplitude and phase of the reflected light when scanning over the chip with a laser.  

    \noindent
    \textbf{Our contributions.}\enskip
    We propose \gls{lit} and \gls{llsi} as techniques for fast and targeted reverse engineering to simplify and speed up following analysis and attacks.
    As a case study, we evaluate our approach on a recent and highly complex technology, i.e., a \gls{soc} manufactured by Intel along their 12\textsubscript{th} Gen. processor series.
    In this regard, we build a custom \gls{pcb} in order to be able to precisely control the individual power rails in an isolated manner.
    Based on our prototype, we show that the position of isolated functional blocks can be identified on the die.
    Finally, we compare \gls{lit} and \gls{llsi} concerning their reverse engineering capabilities, resolution, and acquisition time.

%% file: 03-background/background.tex
\glsreset{llsi}\glsreset{lit}\glsreset{fa}
\section{Background}
    \Gls{fa} represents one of the last steps of the overall \gls{asic} manufacturing process.
    After a wafer of \glspl{ic} has been manufactured by the semiconductor fabrication plant, the so-called yield determines the ratio of functional and non-functional \glspl{ic}.
    For a semiconductor product to be profitable and manufacturers to remain competitive, the yield must be maximized at all costs.
    However, the semiconductor manufacturing process of advanced \glspl{ic} is tremendously complex, i.e., not every part of the process can be controlled in its entirety.
    While this so-called process variation may be utilized positively to build intrinsic \glspl{puf}~\cite{physically-unclonable-functions}, it also implies that a certain percentage of manufactured silicon malfunctions once the variation exceeds a given threshold.

    \noindent
    \gls{fa} is centered around localizing and characterizing a single \gls{ic}'s malfunctioning to tweak future production parameters, thus increasing the yield of future production runs.
    A variety of \gls{fa} techniques exists, each exhibiting advantages and disadvantages in localizing or characterizing a specific kind of fault.
    In this work, we utilize two such \gls{fa} methods, namely \gls{lit}, which is based on thermal principles, and \gls{llsi}, which is based on laser scanning microscopy.
    In this section, both techniques are briefly introduced.
    Moreover, \glspl{pdn} of modern \glspl{soc} are briefly discussed, as they are key to our approach.
    
    \subsection{Thermal Analysis of Integrated Circuits}\label{sec:lit}
		In \gls{fa}, \gls{lit} is employed to localize thermally active regions, which indicate resistive defects in \glspl{ic}.
		Failure analysts use this method to, for example, localize resistive shorts between different metal lines, gate oxide breakdowns, and other faults that cause an increase in contact resistance.
		These resistive defects lead to higher power dissipation and, thus, to a \emph{local} temperature increase.
        As the local temperature increase implies an increase in mid-range \gls{ir} emanation (i.e., $\lambda \in [3..5]$\,\si{\micro\meter}), it can be captured by an \gls{ir}-sensitive camera with high resolution.
		\gls{lit} is based on capturing the thermal radiation in the mid-range \gls{ir} spectrum emitted by an object.

        \noindent
		Resistive defects usually cause power dissipation in the \si{\micro\watt} range, which translates to local temperature differences in the \si{\micro\kelvin} range.
		However, the sensitivity of high-end \gls{ir} sensors lies in the \SI{10}{\milli\watt} range.
		Hence, to be able to measure the small temperature differences, lock-in amplification is mandatory.
		In \gls{lit}, we inject a periodic signal into the \gls{dut}, which is also fed into the lock-in amplifier as a reference.
        The lock-in amplifier then relates the thermal signal captured by the \gls{ir} camera to the reference, filtering and amplifying the thermal information correlating to the induced modulation.

        \noindent
        Moreover, it is worth noting that even a fully powered-off \gls{ic} may exhibit a strong \gls{ir} contrast in emissivity at room temperature due to the difference in emissivity of different materials and structures used in manufacturing.
        Hence, thermographic sensors can be used to record an \gls{ic}'s pattern through the backside.
        
        \begin{figure}[tb]
			\centering
			\includegraphics[width=0.7\linewidth]{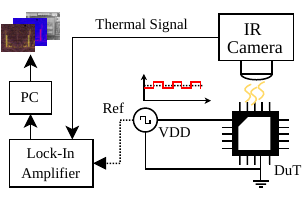}
            \captionsetup{aboveskip=-2pt, belowskip=-8pt}
			\caption{Typical LIT setup.}
			\label{fig:litsetup}
		\end{figure}
        
        \noindent
        Figure~\ref{fig:litsetup} depicts a typical \gls{lit} setup.
        In this work we exclusively consider complex \glspl{soc} exhibiting multiple voltage supplies as \gls{dut}.
        Further, a high-resolution mid-range \gls{ir} camera is required to capture temperature deviations based on a fine scale.
        Different lenses can be used to increase the spatial resolution of the \gls{ir} camera.
        Every \gls{lit} setup requires an external electrical stimulus fed into the \gls{dut}.
        This is commonly achieved using a switchable \gls{psu} that provides the external modulation in the form of a square wave of a given amplitude.
        The lock-in amplifier detects a low-amplitude thermal signal that correlates with the induced signal by performing integration over time.
        Finally, the PC receives temperature amplitude and phase information and stores the results for later analysis.

    \subsection{Laser-Based Analysis of Integrated Circuits}
        Modern \glspl{ic} are comprised of numerous metal layers on the chip's front side, making any analysis through the front side impossible.
        Therefore, analysis is commonly executed through the chip's backside.
        Since silicon is transparent to \gls{nir} light, \glspl{lsm} can be used to access the active area containing the transistors without preparing the silicon backside of the chip.
        One approach to localizing faults is stimulating the \gls{dut} with a laser and measuring the change in resistance, voltage, or current consumption at the device's terminals.
        On the other hand, some part of the laser irradiation is modulated by the electrical characteristics in the chip and reflected at metal interfaces, see \autoref{fig:llsisetup_optical}.
        Consequently, this reflected light contains information about the internal voltages of the chip.
        In \gls{lsm}, a detector captures the reflected light and translates its magnitude and phase into a corresponding signal.
        The approach is part of a family of \gls{fa} methods, referred to as \textit{optical probing} techniques.
        When pointing the laser at one location of interest, a waveform depicting voltage over time can be acquired. The corresponding technique is called \gls{eop}.
        Besides, an activity map can be created when scanning the laser over a larger area of interest and analyzing the reflected light at each point. The technique is called \gls{eofm}, and due to its spatial capabilities, we will focus on \gls{eofm} in the following.

	\subsubsection{Electro-Optical Frequency Mapping}
		\Gls{eofm} is an optical probing technique that allows the creation of a two-dimensional activity map of a circuit area.
		Provided a particular frequency and a bandwidth, \gls{eofm} analyzes the reflected light using a narrow-band frequency filter and maps the resulting amplitude onto the scanning position.
        In this way, all transistors switching at the frequency of interest appear as bright spots in the activity map.
		To not influence the electrical behavior of the \gls{dut}, wavelengths above \SI{1.1}{\micro\meter} are used for optical probing techniques.
        Apart from debugging internal signals in \glspl{ic}, optical probing can be used to attack devices.
        For instance, \gls{eofm} in combination with \gls{eop} has been used to extract sensitive data from a \gls{fpga}~\cite{tajik_power_2017} or to break logic locking schemes~\cite{rahman_key_2020}.

    \subsubsection{Laser-Logic State Imaging}
\begin{figure}[tb]
			\centering
            \subfloat[\label{fig:llsisetup_optical}]{\includegraphics[scale=0.85, trim=0 -.4cm 0 0, clip]{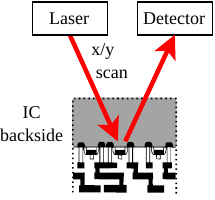}}\hfill
            \subfloat[\label{fig:llsisetup_electrical}]{\includegraphics[scale=0.85]{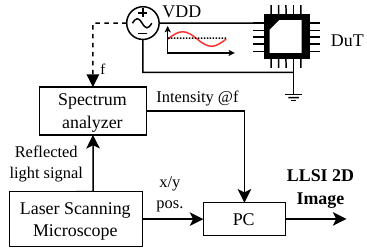}}
            \captionsetup{belowskip=-6pt, aboveskip=4pt}
			\caption{Principle of optical probing (a) and electrical setup for LLSI (b). The supply voltage modulation leads to a detectable pattern in the reflected light, mapped onto the scanning position and shown as a 2D activity map.}
			\label{fig:llsisetup}
		\end{figure}

        \Gls{llsi} is an extension of \gls{eofm} proposed by \Citeauthor{niu_laser_2014}~\cite{niu_laser_2014}.
        Instead of setting the frequency of \gls{eofm} to the frequency of a logic signal generated by the device, a periodic signal is injected into the \gls{dut}'s power supply, as depicted in \autoref{fig:llsisetup_electrical}.
        In other words, the \gls{dut}'s power supply is modulated around the nominal supply voltage with a small peak-to-peak sine signal.
        \gls{eofm} is then used to search for activity based on the introduced modulation frequency.
        Using \gls{llsi}, the logic states of combinatorial and sequential logic can be extracted under the constraint that the clock is stopped for the duration of the measurement~\cite{krachenfel_realworld_2021, krachenfel_automatic_2021}.
        Apart from transistor states, \gls{llsi} measurements reveal the location of capacitive elements, such as decoupling capacitors.
        Consequently, \gls{llsi} can be used to localize circuitry connected to the power supply rail under modulation.

		\subsection{Power Delivery Networks in \gls{asic} Design}
        The \gls{pdn} of an \gls{asic} is responsible for transmitting current from the package pads to the logic blocks and single transistors.
        Its design poses a special difficulty since it is responsible for maintaining a stable voltage during load, voltage fluctuations, and spikes.
        Several other factors, such as the prevention of abrasion effects, overly excessive heat in single spots, and parasitic effects, make the design of \glspl{pdn} a hard task.

        \noindent
        Since modern \glspl{soc} consist of a vast number of different components and all of these components have different characteristics w.r.t. their power consumption, hardware designers decided to supply different components with different physically isolated voltage rails.
        Furthermore, a \gls{soc} might require different voltages, where I/O cells operate at a different voltage level than internal logic cells.
        It is further possible to perform power gating on specific supplies during low power sleep, while only powering the wake-up logic.
        Other reasons might be that only one component on the \gls{soc} consumes excessive power, such as in modern desktop processors, where the high-performance power network is cut off from other maintenance logic.
        All these requirements lead to modern complex \glspl{soc} having complex \glspl{pdn} with multiple voltage rails.

%% file: 04-approach-and-setup/approach-and-setup.tex
\section{Experimental Setup}

    \subsection{Device under Test}
        In order to thoroughly evaluate our novel approach, we decided to utilize a complex, recent-technology \gls{soc} manufactured by Intel, which is referred to as the \gls{pch}~\cite{intel-pch}.
        In the past, an Intel mainboard's chipset was defined by a north bridge and a south bridge, which determined the interconnection between different components.
        The north bridge was handling high-frequency signaling, whereas the south bridge was taking care of lower-frequency communication.
        Due to the constant increase of integration in microelectronics, the north bridge has been integrated into the \gls{cpu} silicon die, whereas the south bridge's functionality as well as other communication protocols (e.g., USB-3 or PCIe) have been merged into another silicon die, referred to as \gls{pch}.
        It is worth noting that Intel's root of trust is a sub-component of the \gls{pch}, whereas, for AMD-based systems, the root of trust is placed within the CPU silicon.
        Because of the high degree of integrated components, Intel's \gls{pch} exposes $12$ physically isolated voltage rails, which need to be supplied by five different voltage levels.
        For saving space and resources, the rails requiring the same voltage level are typically tied together on a \gls{pcb} level whenever possible.
        While this holds true for all commercially available mainboards, tying together the supply of multiple voltage rails prevents isolated modulation.

    \subsection{Custom Printed Circuit Board}
        \begin{figure}[tb]
            \centering
            \includegraphics[width=0.38\textwidth]{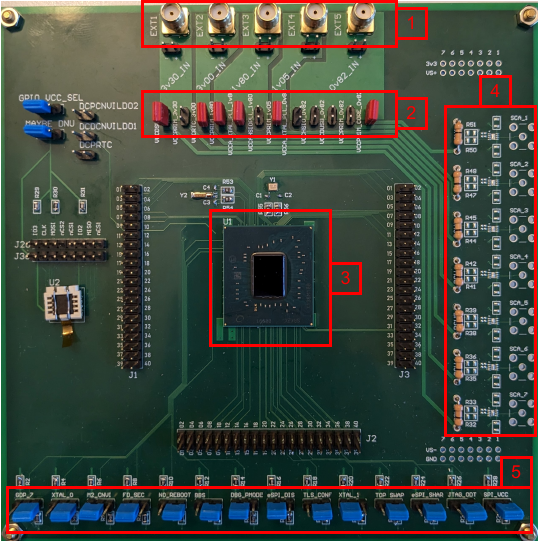}
            \captionsetup{aboveskip=4pt, belowskip=-15pt}
            \caption{\gls{dut} mounted on a custom designed \glsfmtshort{pcb} in order to physically isolate the voltage supplies.}
            \label{fig:pcb}
        \end{figure}
        As the goal of this work is to detect several regions of interest by modulating their supply voltages, we have placed our \gls{dut} on a custom designed \gls{pcb}, which grants us isolated access to each of the voltage rails.
        Our custom \gls{pcb} is depicted in \autoref{fig:pcb}.
        The \gls{pch} must be supplied with $5$ different voltage levels, which are used to supply power to $12$ different, physically isolated voltage rails.
        Different voltage levels may be provided by the SMA connectors \rectangled{1}, a jumper \rectangled{2} then either connects or disconnects a specific voltage rail to the external voltage.
        A set of specific voltage rails has further been connected indirectly via shunt resistors and current sense amplifiers \rectangled{4} to the \gls{dut} \rectangled{3}.
        By this, power-based \gls{sca} attacks are possible for a selected number of voltage rails.
        However, we keep performing \gls{sca} on the different voltage rails of the \gls{pch} as future work.
        Moreover, different boot configurations can be chosen by configuring the jumpers in \rectangled{5}.

    \subsection{Measurement Setup} 
    \subsubsection{\gls{lit} Setup}\label{sec:litsetup}
        The \gls{lit} setup is equal to the one depicted in \autoref{fig:litsetup}.
        The \gls{dut} is represented by Intel's \gls{pch}, which is mounted on our custom \gls{pcb}.
        By exposing each voltage rail in a physically isolated fashion, we are able to modulate each rail without affecting the others.
        Here, the modulation takes place based on a periodic square wave signal in the $40-\SI{60}{\hertz}$ range, which can be generated directly by a software controlled \gls{psu}.
        The silicon die's mid-\gls{ir} emanation in the field of view of the optical lens is sampled by the camera.
        The recorded data is forwarded to the lock-in amplifier, which is also provided with the switched power supply as a reference.
        
    \subsubsection{\gls{llsi} Setup} \label{sec:llsisetup}
        While using the same \gls{dut} (i.e., Intel's $610$ \gls{pch} on custom \gls{pcb}), each voltage rail can be modulated by a much higher frequency than it is possible for \gls{lit}, thus decimating noise.
        As common \glspl{psu} are incapable of providing modulation in the MHz range, a Bias-Tee in combination with a function generator and a DC \gls{psu} have been used to generate a \SI{2}{\mega\hertz} sine-modulated voltage supply signal.
        For conducting the \gls{llsi} measurements, we use a Hamamatsu PHEMOS-1000 \gls{fa} microscope, which offers lenses of $5\times$, $20\times$, and $50\times$ magnification.

%% file: 05-results/results.tex
\section{Evaluation}
    In this section, we showcase the effectiveness of our technique.
    By modulating different voltage rails of the \gls{pch} utilizing our \gls{pcb} design, we can clearly distinguish between different regions.
    We present the results of two different measurements, namely modulating \texttt{vcc\_core\_0p82} as well as \texttt{vcc\_usb\_0p82}\footnote{Following Intel's nomenclature: \url{https://www.intel.com/content/www/us/en/products/sku/218829/intel-h610-chipset/specifications.html}}.
    We have selected these two scenarios as a representative subset, as they highlight the different outcomes of our measurements.

    \noindent
    As a metric to quantify the reduction in search space achieved by our technique, we compute the area that responds to the external modulation.
    The evaluation takes place based on thresholding, i.e., if a signal within a region exceeds a threshold, we classify it as being affected by our modulation, otherwise, it is classified as unaffected.
    As without modulation an adversary is required to scan the entire die, we compare the die's overall area to our identified regions to quantify the area reduction using our technique.

    \noindent
    The minimum time a physical attack with spatial information requires can be approximated by considering the number of positions to be tested, the time per attempt, and the number of attempts per position.
    In addition to the aforementioned parameters, when considering \gls{fi} also all combinations of the fault's parameters have to be considered (e.g., offset and strength).
    \begin{equation*}
        t_{\text{scan}} =
        \frac{\text{area width}}{\text{step size x}} \times \frac{\text{area height}}{\text{step size y}} \times n \times t_{\text{attempt}} \times \text{comb}_{\text{params}}
    \end{equation*}

    \noindent
    As an example, when considering \gls{lfi}, a magnification of $50\times$ is commonly required to induce enough energy within a spatially limited radius for the photoelectric effect to cause logical misbehavior at the transistor level.
    A $50\times$ lens commonly corresponds to a transistor-focused laser spot size of about \SI{1}{\micro\meter}.
    Hence, a step size in either \texttt{x} or \texttt{y} of \SI{1}{\micro\meter} must not be exceeded.
    In our case study, the silicon die is \SI{8}{\milli\meter} wide and \SI{12}{\milli\meter} high, which -- based on a step size of \SI{1}{\micro\meter} in \texttt{x} and \texttt{y} -- results in $96,000,000$ possible positions.
    Even when considering a single attempt per position ($n=1$), a single combination of fault parameters ($\text{comb}_{\text{params}}=1$) and a time per attempt of \SI{0.1 }{\second} ($t_{\text{attempt}} = 0.1$), $111$ days would be required to scan the whole die area.
    For this simplified approximation, the time required to move the stage and re-focus the laser along the Z-axis is neglected.
    
    \subsection{Modulation of \texttt{vcc\_core\_prim\_0p82}}
        As the name implies, \texttt{vcc\_core\_prim\_0p82} appears to power the primary core logic contained inside the \gls{pch}, whereas \texttt{0p82} indicates an electrical potential of \SI{0.82}{\volt}.
        In the following, we present the results of performing \gls{lit} as well as \gls{llsi} based on a modulation of \texttt{vcc\_core\_prim\_0p82}.
        As the hereby identified regions represent core logic components, these form potentially lucrative areas for further physical attacks.
        \begin{figure*}[t!]
            \centering
            \subfloat[\gls{lit}, captured with $1\times$ magnification.\label{fig:vcore_prim_lit}]{
                \begin{tikzpicture}
                    \node[inner sep=0, anchor=south west] (pic) {\includegraphics[height=.33\linewidth, trim=0 1.5cm 0 1cm, clip]{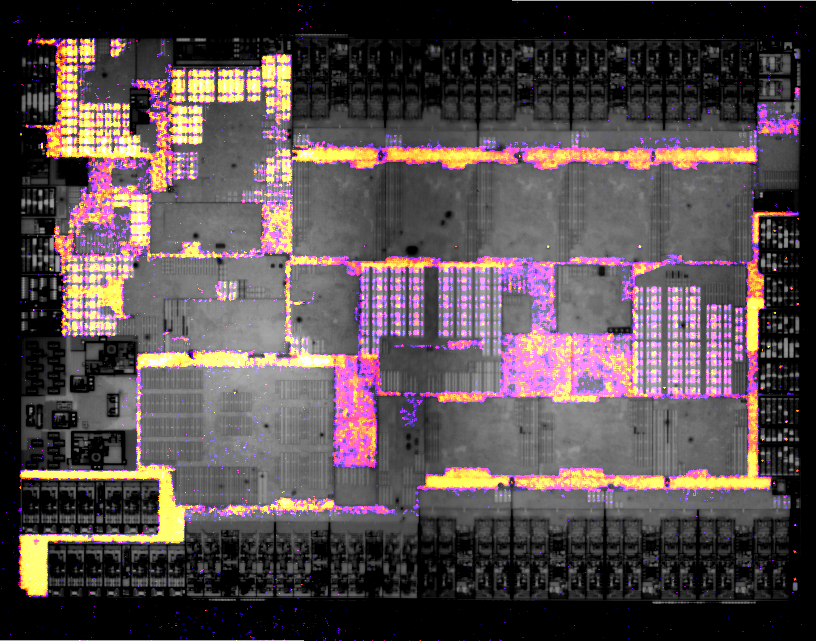}};
                    \scalebarbackgroundtrconfunit{pic}{817}{463/5118*0.781679389312977*1000}{2}{\si{\milli\meter}};
                \end{tikzpicture}
            }\hfill
            \subfloat[\gls{llsi}, captured with 20$\times$ magnification (stitched from 204 images)\label{fig:vcore_prim_llsi}]{
                \begin{tikzpicture}
                    \node[inner sep=0, anchor=south west] (pic) {\includegraphics[height=.33\linewidth]{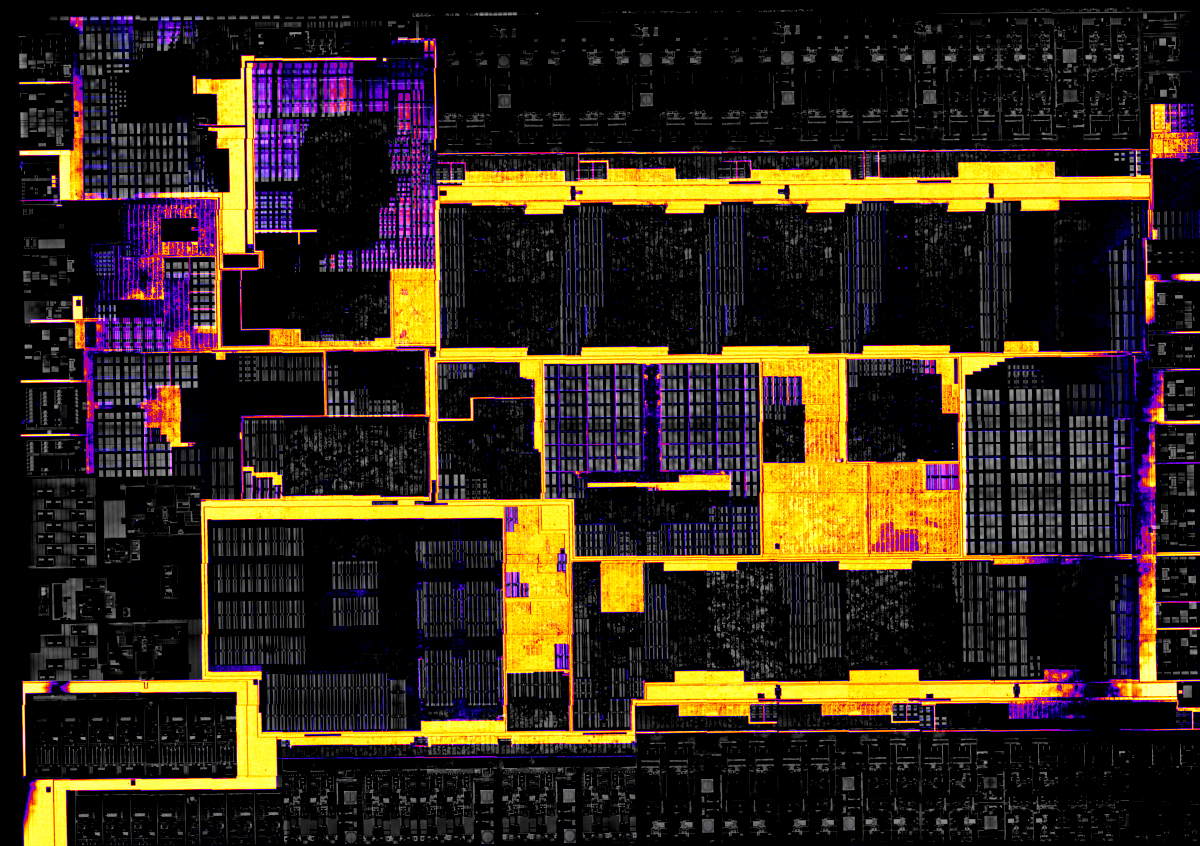}};
                    \scalebarbackgroundtrconfunit{pic}{1200}{1200/8678*0.781679389312977*1000}{2}{\si{\milli\meter}};
                \end{tikzpicture}
            }
            \label{fig:vcore_prim}
            \captionsetup{aboveskip=4pt}
            \caption{\gls{lit} and \gls{llsi} amplitudes overlaid on the optical image for the \texttt{vcc\_core\_prim\_0p82} rail.}
        \end{figure*}

        \subsubsection{\gls{lit}}
            The results of modulating \texttt{vcc\_core\_prim\_0p82} and performing \gls{lit} as described in \autoref{sec:litsetup} are depicted in \autoref{fig:vcore_prim_lit}.
            Here, a yellow overlay indicates that after the \gls{lit} process, a strong increase in temperature was recognized in the corresponding region, whereas purple indicates, that minor temperature deviation has been noted which matches the induced modulation frequency.
            The remaining regions are completely unaffected by the external modulation.
            By modulating \texttt{vcc\_core\_prim\_0p82}, we obtained a \gls{lit} signal that covers about \SI{18.9}{\percent} of the chip area.
            This corresponds to a search space reduction of \SI{81.1}{\percent} compared to an exhaustive scan.
            However, in order to even further narrow down the search space, we continue to analyze the different emissivity characteristics of different structures. 
            As depicted in the thermal image, different areas of different intensity values were captured.
            While the solid yellow areas, where the highest intensity is observed, can be expected to belong to power supply circuitry (i.e., \gls{pdn} structures), the yellow-purple sprinkled areas are promising candidates for synthesized logic cores.
            The difference is depicted in more detail in \autoref{fig:vcore_prim_lit_crop}.
            Using bare eyes, the remaining search space can therefore be cut again, leading to a potential target chip area of only \SI{15.4}{\percent}.
        
        \subsubsection{\gls{llsi}}
\begin{figure}[t!]
            \centering
            \subfloat[LIT\label{fig:vcore_prim_lit_crop}]{
            \begin{tikzpicture}
                \node[inner sep=0, anchor=south west] (pic) {\includegraphics[height=.35\linewidth, trim=0 0 2cm 2cm, clip]{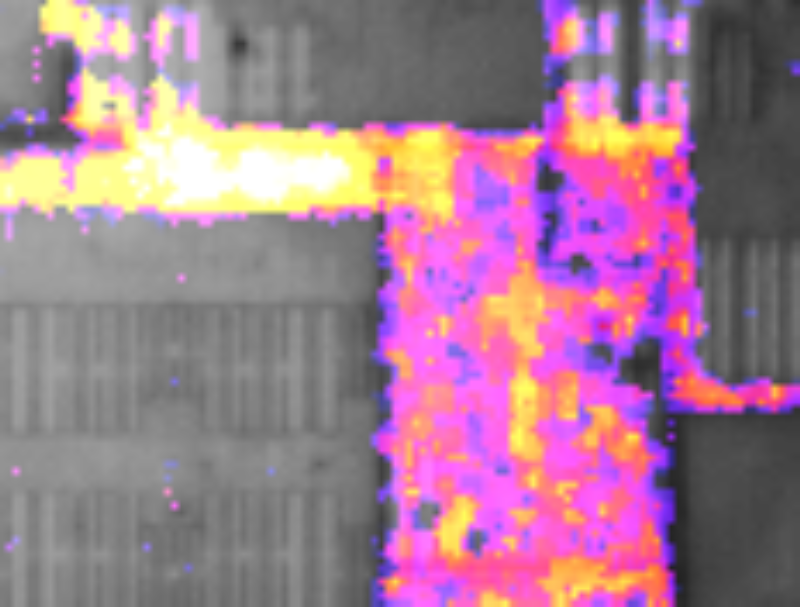}};
                \scalebarbackgroundtrconf{pic}{800}{0.090465025*7.5*0.781679389312977}{400};

                \node[inner sep=1mm, anchor=west, fill=white, fill opacity=0.7, text opacity=1] at (img cs:node=pic,x=0.1,y=0.5) (supp) {\small Supply};
                \draw[->, red, ultra thick, dotted] (supp.north) -- (img cs:node=pic,x=0.25,y=0.79);

                \node[inner sep=1mm, anchor=west, fill=white, fill opacity=0.7, text opacity=1] at (img cs:node=pic,x=0.1,y=0.2) (log) {\small Logic};
                \draw[->, red, ultra thick, dotted] (log.east) -- (img cs:node=pic,x=0.59,y=0.3);
            \end{tikzpicture}
            }\hfill
            \subfloat[LLSI\label{fig:vcore_prim_llsi_crop}]{
            \begin{tikzpicture}
                \node[inner sep=0, anchor=south west] (pic) {\includegraphics[height=.35\linewidth]{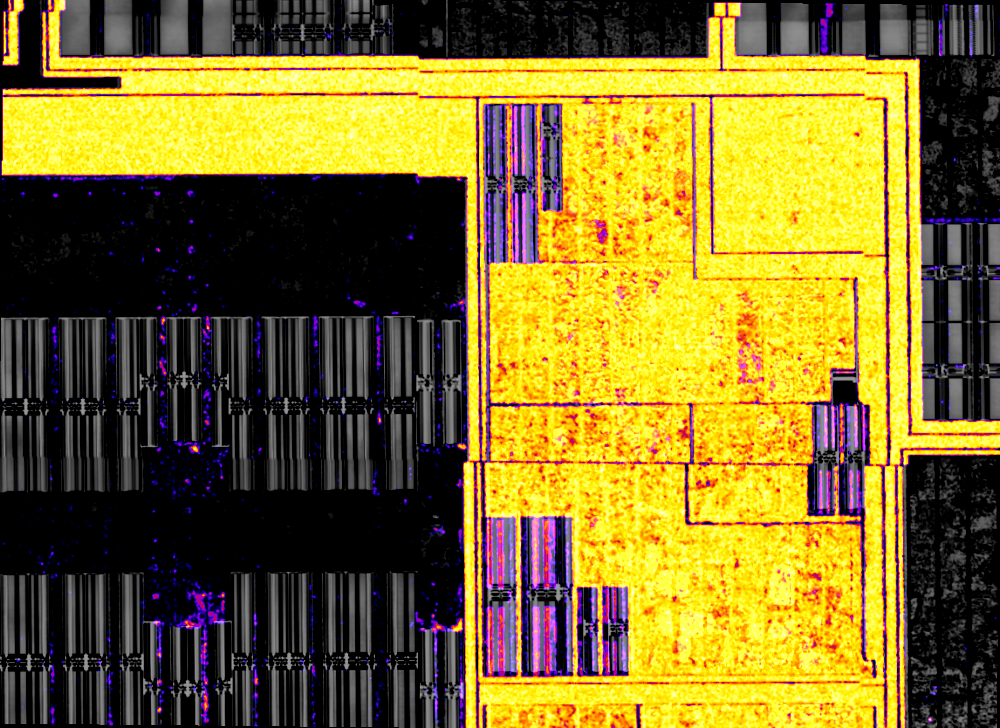}};
            \end{tikzpicture}
            }
            \captionsetup{aboveskip=4pt, belowskip=-12pt}
            \caption{Comparison of \gls{lit} and \gls{llsi} in one region of interest to show the possibility of distinguishing between power supply and logic areas.\label{vcore_prim_zoom}}
            \label{fig:logic-vs-pdn}
        \end{figure}
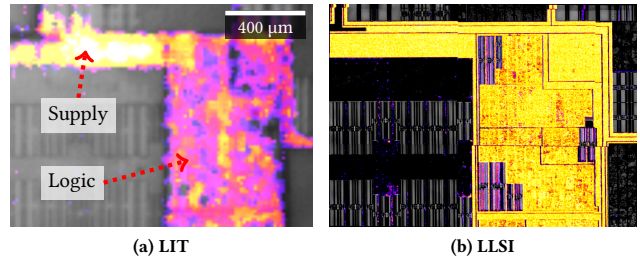

            The results of modulating \texttt{vcc\_core\_prim\_0p82} and scanning over the die as described in \autoref{sec:llsisetup} are depicted in \autoref{fig:vcore_prim_llsi}.
            Again, yellow indicates that the modulation in the reflected light shows a strong correlation in amplitude with our injected stimulus, whereas purple indicates, that the modulation of the reflected light slightly diminishes.
            All remaining regions are not affected by modulation at all.
            It is worth noting that the regions appearing speckled in the \gls{lit} measurements show up as speckled again.
            However, the regions identified by \gls{llsi} as well as the speckle pattern are much more precise and sharp.
            Their difference in the same region as before is depicted in \autoref{fig:vcore_prim_llsi_crop}.
            By modulating \texttt{vcc\_core\_prim\_0p82}, we obtained an \gls{llsi} signal that covers about \SI{16.3}{\percent} of the chip area, i.e., \SI{2.6}{\percent} less area than measured by \gls{lit}.
            This corresponds to a search space reduction of \SI{83.7}{\percent} compared to an exhaustive scan.
            Same as before, by considering the differences of solid \gls{pdn} area and speckled logic area, this time the search space can even be reduced to \SI{10.9}{\percent}, i.e., \SI{4.5}{\percent} less than with \gls{lit}.

        \subsection{Modulation of \texttt{vcc\_usb\_0p82}}
            While the previous measurement revealed that \gls{lit} and \gls{llsi} are both capable of identifying \glspl{pdn} as well as their supplied logic, with this experiment we would like to show that these techniques can also be used to uniquely identify regions that are right next to each other without any interference.
            The \texttt{vcc\_usb\_0p82} appears to power the USB logic contained inside the \gls{pch}, whereas \texttt{0p82} indicates an electrical potential of \SI{0.82}{\volt}.
            In the following, we present the results of performing \gls{lit} by modulating \texttt{vcc\_usb\_0p82}.
            While the results of performing \gls{llsi} are similar, they have been omitted due to space constraints. 
            However, high-resolution images of applying \gls{llsi} are provided in the appendix in \autoref{fig:appendix_vccusb_llsi_zoom}.
        
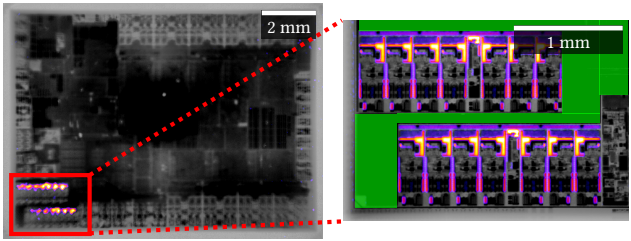
\begin{figure}[h!]
                    \centering
                    \import{images/results/LIT/USB/}{USB_zoom-in.tex}
                    \captionsetup{belowskip=-12pt, aboveskip=4pt}
                    \caption{LIT amplitude for the \texttt{vcc\_usb\_0p82} rail captured with 1$\times$ (left) and 2.5$\times$ (right) magnification.
                    The adjacent regions previously identified to belong to \texttt{vcc\_core\_prim\_0p82} are depicted in green.
                    }
                    \label{fig:vcc_usb_lit}
                \end{figure}

                \noindent
                The results of modulating the \texttt{vcc\_usb\_0p82} voltage and performing \gls{lit} as described in \autoref{sec:litsetup} are depicted in \autoref{fig:vcc_usb_lit}.
                It is important to note that compared to the previous measurement, a relatively small area of the die shows a thermal correlation to the modulation.
                As before, a yellow overlay indicates a strong increment in local \gls{ir} emissivity correlating to the modulation, whereas purple indicates a weaker emissivity.
                All other regions are unaffected by the external modulation of \texttt{vcc\_usb\_0p82}.
                By modulating the USB supply voltage, we successfully identified this part of the \gls{soc}, which handles the USB protocol communication. It only covers \SI{1.2}{\percent} of the die area.
                Moreover, when superimposing the results of the previous measurement (i.e., the modulation of \texttt{vcc\_core\_prim\_0p82}), the proximity of the results becomes observable.
                By this, we provide proof that our technique for reverse engineering can be used with high spatial resolution.

%% file: images/results/LIT/USB/USB_zoom-in.tex
\begin{tikzpicture}
    \node[inner sep=0, anchor=south west] (pic) {\includegraphics[width=.5\linewidth]{Composite_MACRO.png}};
    \scalebarbackgroundtrconfunit{pic}{518}{287/5118*0.781679389312977*1000}{2}{\si{\milli\meter}};

    \node[inner sep=0, anchor=south west, right=.25cm of pic.east] (pic-zoom) {\includegraphics[width=.45\linewidth]{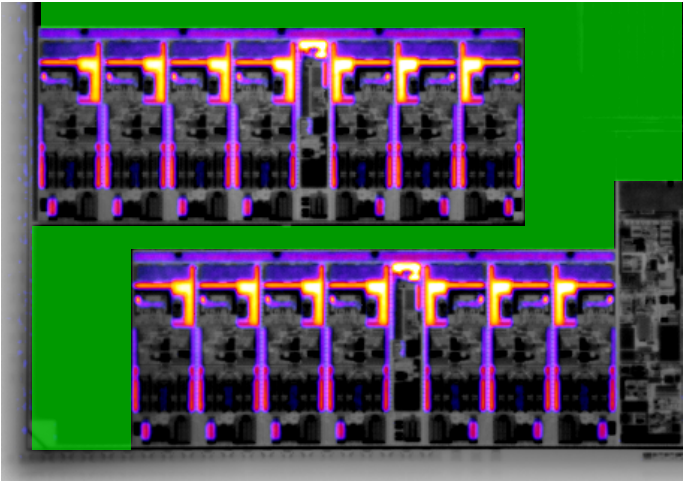}};
    \scalebarbackgroundtrconfunit{pic-zoom}{456}{328/1488*0.781679389312977*1000}{1}{\si{\milli\meter}};

    \draw[red, ultra thick] (img cs:node=pic,x=0.025,y=0.02) rectangle (img cs:node=pic,x=0.27,y=0.27);
    \draw[red, ultra thick, dotted] (img cs:node=pic,x=0.27,y=0.27) -- (img cs:node=pic-zoom,x=0,y=0.991);
    \draw[red, ultra thick, dotted] (img cs:node=pic,x=0.27,y=0.023) -- (img cs:node=pic-zoom,x=0,y=0.01);
\end{tikzpicture}

%% file: 06-discussion/discussion.tex
\section{Discussion}
    In this work, we have utilized \gls{lit} and \gls{llsi} to discover the position of a specific circuitry of our target.
    For both setups, we provided external modulation of a given frequency to discover regions connected to physically isolated \glspl{pdn}.
    Since \gls{lit} and \gls{llsi} exhibit similar capabilities and results during our evaluation, we discuss the main differences between both techniques before concluding this work.
    
    \subsection{Spatial Resolution and Acquisition Time}
    In this work, the spatial resolution of \gls{llsi} was much higher than this of \gls{lit}.
    This is due to the fact that for \gls{lit}, commonly only low-magnification lenses with sufficiently good optical properties are available.
    Due to the poor properties of the lenses, a higher magnification drastically increases the measurement time.
    During our measurements, only weak signals have been recorded with lenses of $10\times$ magnification.
    Nevertheless, the \gls{lit} images presented in this work, captured with a 1$\times$ lens, could compete with the results obtained by applying \gls{llsi}.
    Vice-versa, \gls{llsi} measurements with a reasonable signal-to-noise ratio could only be obtained with the 20$\times$ lens and above, making the scan comparably slow.
    While for \gls{lit} the scanning time was in the range of a few hours for the entire chip, scanning the die in an automated fashion using \gls{llsi} with the 20$\times$ lens took roughly one day.
    Consequently, for a first overview, \gls{lit} can deliver sufficient and fast results.
    When higher magnification for more detailed analysis is required, \gls{llsi} should be considered.

    \subsection{Setup Cost and Availability}
    While the \gls{lit} setup used in this work can be acquired for around \$200K, a setup for optical probing costs at least \$1M.
    Consequently, \gls{lit} can be considered the more cost-efficient solution.
    However, there is always the possibility to rent \gls{fa} equipment or even to hire a failure analyst in a much more affordable way.
    
    \subsection{Backside Silicon Access}
    Direct access to the silicon surface is a strict requirement for optical probing methods.
    Moreover, the silicon substrate must fulfill specific properties (e.g., polished surface, no highly-doped silicon).
    Although flip-chip packages have become more relevant over the past years, less complex \glspl{ic} are still packaged by other means, which often encapsulate the \gls{ic} in a plastics or ceramic case.
    Hence, to perform optical inspection, the \gls{ic} has to be decapsulated and polished, which is a tedious and risky process, as it may result in a broken \gls{dut}.
    Methods used range from chemical to mechanical processes, and each step must be taken carefully to leave the device operable after decapsulation.

    \noindent
    In this regard, \gls{lit} has an advantage over \gls{llsi}: it is a \gls{fa} method that does not strictly require the silicon backside to be exposed.
    \gls{lit} measurements are typically also possible through a package, though the spatial resolution decreases when compared to  silicon is accessible.
    We expect that \gls{lit} delivers results that are acceptable for \gls{emfi}, as it is a less location-dependent physical attack than, e.g., \gls{lfi}.
    Although we did not perform experiments with this scenario, it is an intriguing approach for further investigation.

%% file: 07-conclusion/conclusion.tex
\section{Conclusion}
In this paper, we presented a novel method leveraging \gls{lit} and \gls{llsi} to identify specific parts of circuitry on a large, fully unknown \gls{soc}.
Advanced high-performance \glspl{ic} always expose multiple voltage rails, which provide the power to different sub-circuitry.
The modulation of different voltage supplies allows optical as well as thermal techniques to map a voltage rail to specific regions that are powered by the corresponding supply.
As voltage rails commonly need to be labeled, an adversary may deduce semantic information about the identified circuitry.
While not focusing on introducing a specific attack, we provide a building block that makes physical attacks requiring spatial information even feasible.

\noindent
Moreover, we have provided proof that our method works well for a recent-technology Intel \gls{pch}, where we were able to identify subcircuits with ease.
By using our novel approach, it was possible to identify the exact positions and sizes of USB, RTC, and core logic, thus drastically reducing the search space of a subsequent attack.

%% file: 08-appendix/appendix.tex
\onecolumn
\section{Appendix}

\begin{figure}[H]
    \centering
    \begin{tikzpicture}
        \node[inner sep=0, anchor=south west] (pic) {\includegraphics[width=.7\linewidth]{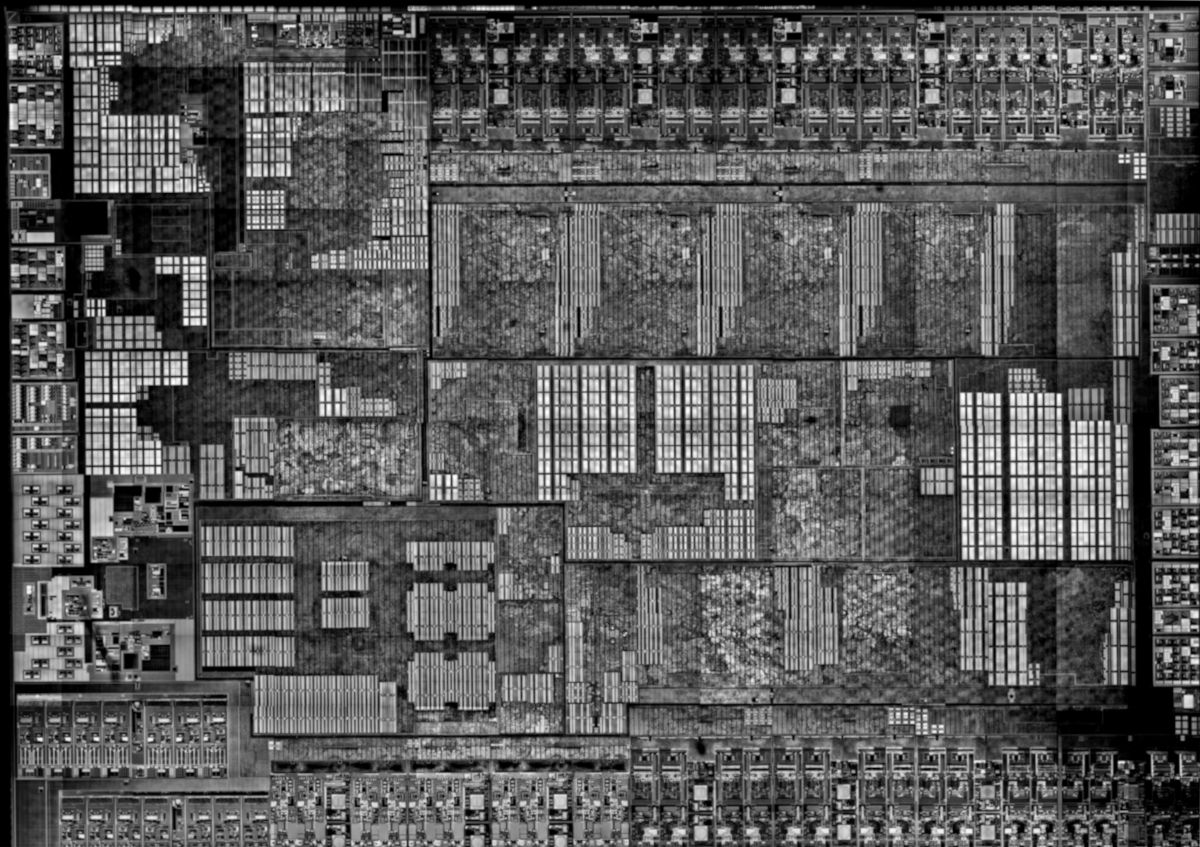}};
        \scalebarbackgroundtrconfunit{pic}{1200}{1200/8678*0.781679389312977*1000}{2}{\si{\milli\meter}};
    \end{tikzpicture}
    
    \caption{Reflected light laser scanning image stitched from 204 images captured with the 20$\times$ lens.}
    \label{fig:appendix_LLSI_pattern}
\end{figure}

\begin{figure}[H]
    \centering
    \begin{tikzpicture}
        \node[inner sep=0, anchor=south west] (pic) {\includegraphics[width=.7\linewidth]{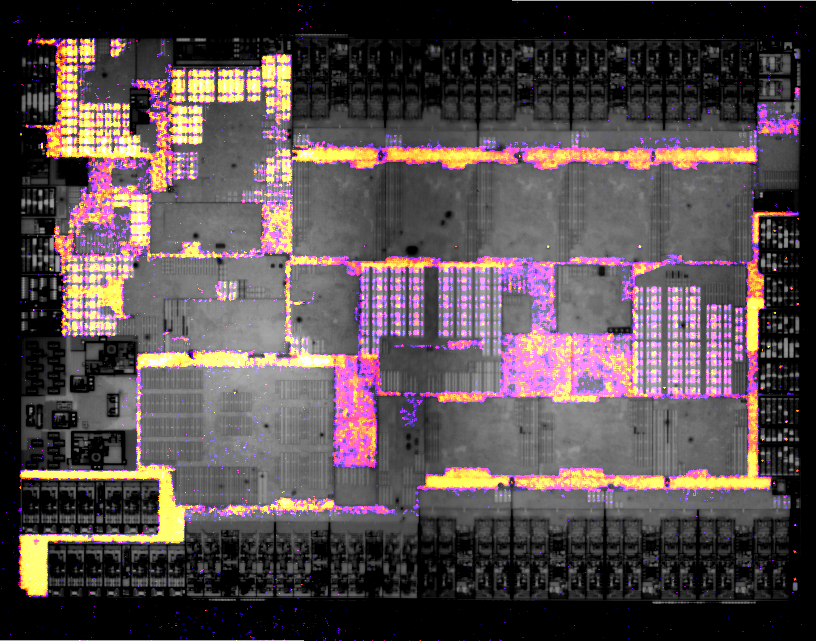}};
        \scalebarbackgroundtrconfunit{pic}{817}{463/5118*0.781679389312977*1000}{2}{\si{\milli\meter}};
    \end{tikzpicture}
    \caption{\gls{lit} amplitude overlaid on the optical image for the \texttt{vcc\_core\_prim\_0p82} rail captured with $1\times$ magnification.}
    \label{fig:appendix_vcore_prim_lit}
\end{figure}

\begin{figure}[H]
    \centering
    \begin{tikzpicture}
        \node[inner sep=0, anchor=south west] (pic) {\includegraphics[width=.7\linewidth]{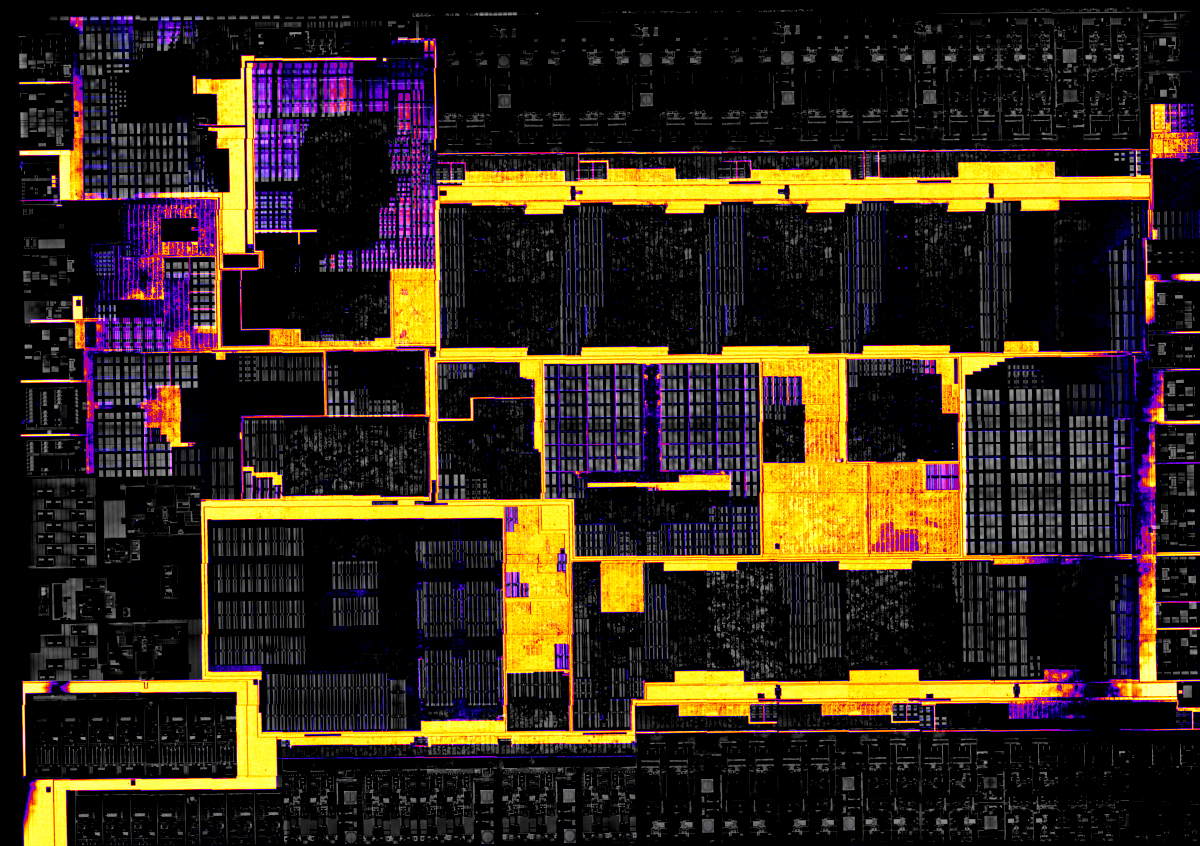}};
        \scalebarbackgroundtrconfunit{pic}{1200}{1200/8678*0.781679389312977*1000}{2}{\si{\milli\meter}};
    \end{tikzpicture}
    \caption{\gls{llsi} amplitude overlaid on the optical image for the \texttt{vcc\_core\_prim\_0p82} rail captured with $20\times$ magnification.}
    \label{fig:appendix-vcoreprim_llsi}
\end{figure}

\begin{figure}[H]
    \centering
    \begin{tikzpicture}
    \node[inner sep=0, anchor=south west] (pic) {\includegraphics[width=.7\linewidth]{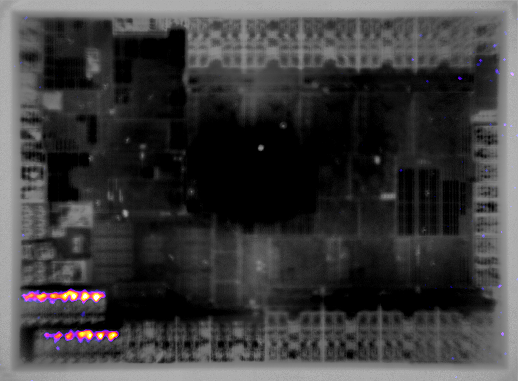}};
    \scalebarbackgroundtrconfunit{pic}{518}{287/5118*0.781679389312977*1000}{2}{\si{\milli\meter}};
\end{tikzpicture}
    \caption{LIT amplitude overlaid on the optical image for the \texttt{vcc\_usb\_0p82} rail captured with the macro lens.}
    \label{fig:appendix_vccusb_lit}
\end{figure}


\begin{figure}[H]
    \centering
    \begin{tikzpicture}
    \node[inner sep=0, anchor=south west] (pic) {\includegraphics[width=.7\linewidth]{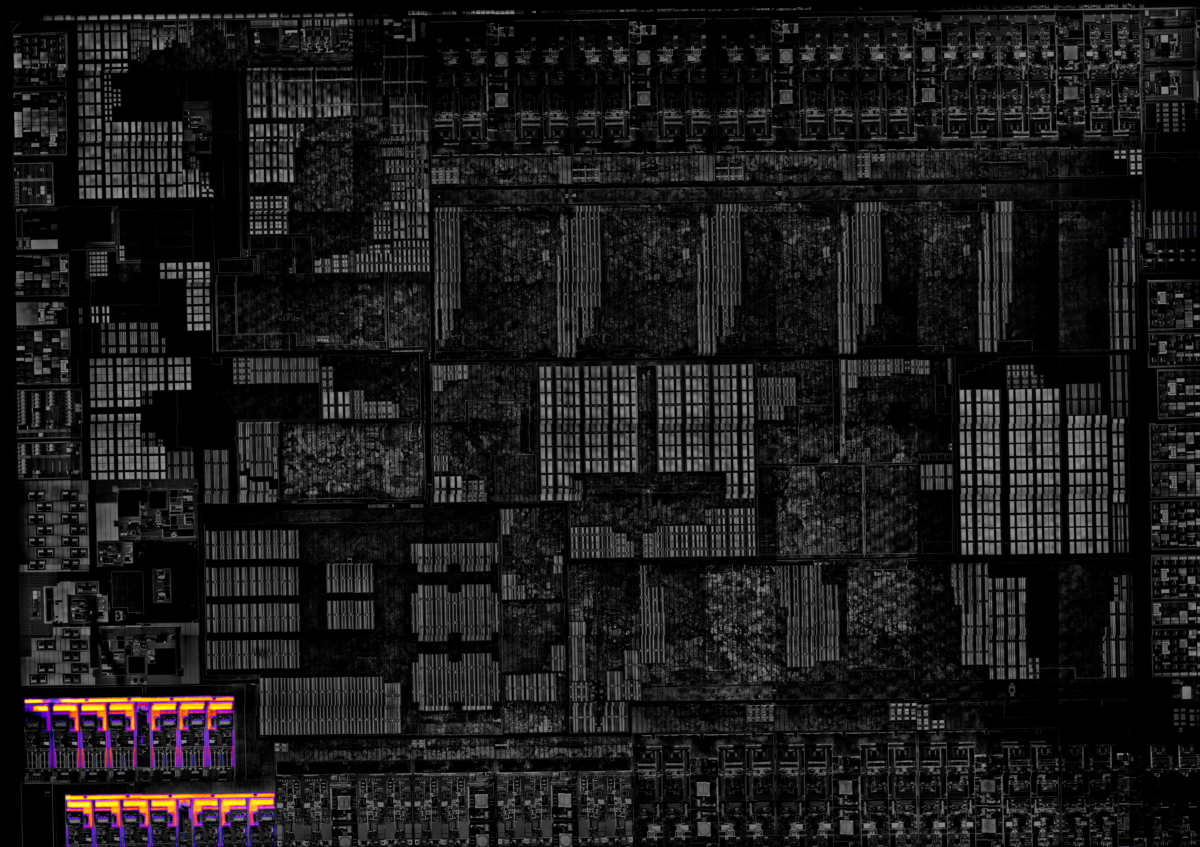}};
    \scalebarbackgroundtrconfunit{pic}{1200}{1200/8704*0.781679389312977*1000}{2}{\si{\milli\meter}};
    \end{tikzpicture}
    \caption{\gls{llsi} amplitude overlaid on the optical image for the \texttt{vcc\_usb\_0p82} rail captured with the $20\times$ lens.}
    \label{fig:appendix-vccusb_llsi}
\end{figure}

\begin{figure}[H]
    \centering
    \import{images/results/LLSI/USB/}{USB_zoom-in.tex}
    \caption{LLSI amplitude overlaid on the optical image for the \texttt{vcc\_usb\_0p82} rail captured with 20$\times$ magnification.}
    \label{fig:appendix_vccusb_llsi_zoom}
\end{figure}
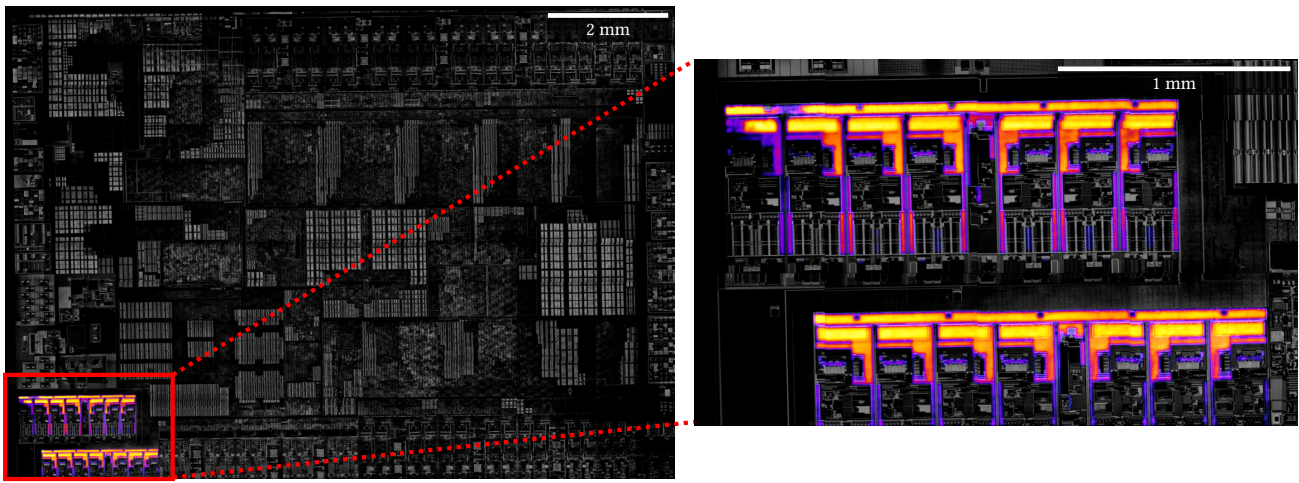

%% file: images/results/LLSI/USB/USB_zoom-in.tex
\begin{tikzpicture}
    \node[inner sep=0, anchor=south west] (pic) {\includegraphics[width=.5\linewidth]{Composite_PATTERN_AMP1.png}};
    \scalebarbackgroundtrconfunit{pic}{1200}{1200/8704*0.781679389312977*1000}{2}{\si{\milli\meter}};

    \node[inner sep=0, anchor=south west, right=.25cm of pic.east] (pic-zoom) {\includegraphics[width=.45\linewidth]{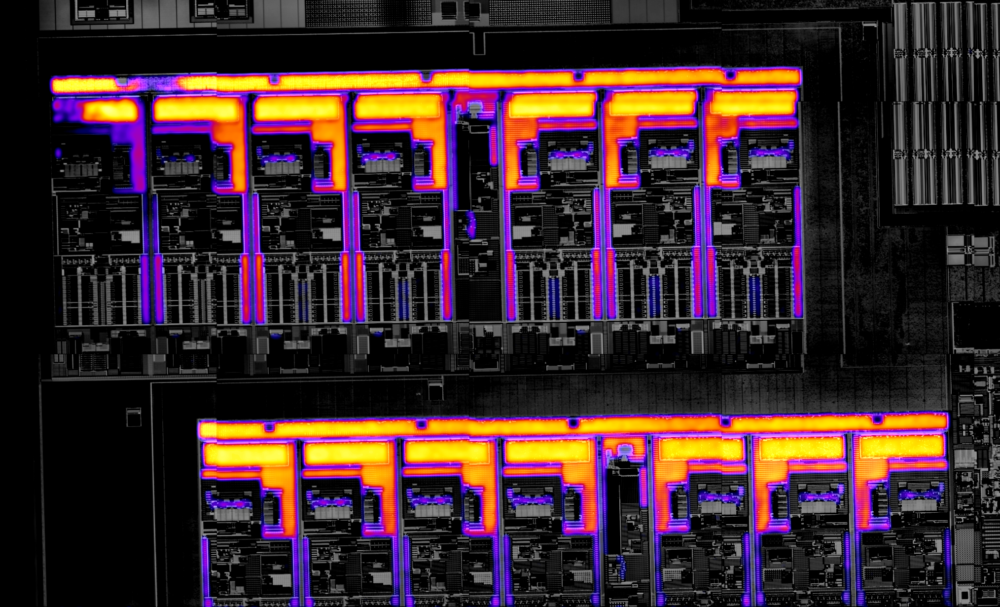}};
    \scalebarbackgroundtrconfunit{pic-zoom}{1000}{1000/2028*0.781679389312977*1000}{1}{\si{\milli\meter}};

    \draw[red, ultra thick] (img cs:node=pic,x=0.0,y=0.0) rectangle (img cs:node=pic,x=0.25,y=0.22);
    \draw[red, ultra thick, dotted] (img cs:node=pic,x=0.25,y=0.22) -- (img cs:node=pic-zoom,x=0,y=0.997);
    \draw[red, ultra thick, dotted] (img cs:node=pic,x=0.25,y=0.003) -- (img cs:node=pic-zoom,x=0,y=0.01);
\end{tikzpicture}

%% file: main.bbl

\begin{thebibliography}{13}


\ifx \showCODEN    \undefined \def \showCODEN     #1{\unskip}     \fi
\ifx \showDOI      \undefined \def \showDOI       #1{#1}\fi
\ifx \showISBNx    \undefined \def \showISBNx     #1{\unskip}     \fi
\ifx \showISBNxiii \undefined \def \showISBNxiii  #1{\unskip}     \fi
\ifx \showISSN     \undefined \def \showISSN      #1{\unskip}     \fi
\ifx \showLCCN     \undefined \def \showLCCN      #1{\unskip}     \fi
\ifx \shownote     \undefined \def \shownote      #1{#1}          \fi
\ifx \showarticletitle \undefined \def \showarticletitle #1{#1}   \fi
\ifx \showURL      \undefined \def \showURL       {\relax}        \fi
\providecommand\bibfield[2]{#2}
\providecommand\bibinfo[2]{#2}
\providecommand\natexlab[1]{#1}
\providecommand\showeprint[2][]{arXiv:#2}

\bibitem[\protect\citeauthoryear{Intel}{Intel}{2022}]%
        {intel-pch}
Intel \bibinfo{year}{2022}\natexlab{}.
\newblock \bibinfo{booktitle}{\emph{Intel 600 Series Chipset Family Platform
  Controller Hub}}.
\newblock Intel.
\newblock
\newblock
\shownote{Rev. 004.}


\bibitem[\protect\citeauthoryear{Krachenfels, Ganji, Moradi, Tajik, and
  Seifert}{Krachenfels et~al\mbox{.}}{2021a}]%
        {krachenfel_realworld_2021}
\bibfield{author}{\bibinfo{person}{Thilo Krachenfels}, \bibinfo{person}{Fatemeh
  Ganji}, \bibinfo{person}{Amir Moradi}, \bibinfo{person}{Shahin Tajik}, {and}
  \bibinfo{person}{Jean-Pierre Seifert}.} \bibinfo{year}{2021}\natexlab{a}.
\newblock \showarticletitle{Real-{{World Snapshots}} vs. {{Theory}}:
  {{Questioning}} the t-{{Probing Security Model}}}. In
  \bibinfo{booktitle}{\emph{2021 {{IEEE Symposium}} on {{Security}} and
  {{Privacy}} ({{SP}})}}. \bibinfo{publisher}{{IEEE Computer Society}},
  \bibinfo{pages}{1955--1971}.
\newblock
\urldef\tempurl%
\url{https://doi.org/10.1109/SP40001.2021.00029}
\showDOI{\tempurl}


\bibitem[\protect\citeauthoryear{Krachenfels, Kiyan, Tajik, and
  Seifert}{Krachenfels et~al\mbox{.}}{2021b}]%
        {krachenfel_automatic_2021}
\bibfield{author}{\bibinfo{person}{Thilo Krachenfels}, \bibinfo{person}{Tuba
  Kiyan}, \bibinfo{person}{Shahin Tajik}, {and} \bibinfo{person}{Jean-Pierre
  Seifert}.} \bibinfo{year}{2021}\natexlab{b}.
\newblock \showarticletitle{Automatic {{Extraction}} of {{Secrets}} from the
  {{Transistor Jungle}} using {{Laser-Assisted Side-Channel Attacks}}}. In
  \bibinfo{booktitle}{\emph{30th {{USENIX Security Symposium}} ({{USENIX
  Security}} 21)}}. \bibinfo{publisher}{{USENIX Association}},
  \bibinfo{pages}{627--644}.
\newblock
\showISBNx{978-1-939133-24-3}


\bibitem[\protect\citeauthoryear{Kühnapfel, Buhren, Jacob, Krachenfels,
  Werling, and Seifert}{Kühnapfel et~al\mbox{.}}{2022}]%
        {AMD_EM_fault}
\bibfield{author}{\bibinfo{person}{Niclas Kühnapfel}, \bibinfo{person}{Robert
  Buhren}, \bibinfo{person}{Hans~Niklas Jacob}, \bibinfo{person}{Thilo
  Krachenfels}, \bibinfo{person}{Christian Werling}, {and}
  \bibinfo{person}{Jean-Pierre Seifert}.} \bibinfo{year}{2022}\natexlab{}.
\newblock \showarticletitle{EM-Fault It Yourself: Building a Replicable EMFI
  Setup for Desktop and Server Hardware}. In \bibinfo{booktitle}{\emph{2022
  IEEE Physical Assurance and Inspection of Electronics (PAINE)}}.
  \bibinfo{pages}{1--7}.
\newblock
\urldef\tempurl%
\url{https://doi.org/10.1109/PAINE56030.2022.10014927}
\showDOI{\tempurl}


\bibitem[\protect\citeauthoryear{Niu, Khoo, Chen, Chapman, Bockelman, and
  Tong}{Niu et~al\mbox{.}}{2014}]%
        {niu_laser_2014}
\bibfield{author}{\bibinfo{person}{Baohua Niu}, \bibinfo{person}{Grace Mei~Ee
  Khoo}, \bibinfo{person}{Yuan-Chuan~Steven Chen}, \bibinfo{person}{Fernando
  Chapman}, \bibinfo{person}{Dan Bockelman}, {and} \bibinfo{person}{Tom Tong}.}
  \bibinfo{year}{2014}\natexlab{}.
\newblock \showarticletitle{Laser {{Logic State Imaging}} ({{LLSI}})}. In
  \bibinfo{booktitle}{\emph{{{ISTFA}} 2014: {{Conference Proceedings}} from the
  40th {{International Symposium}} for {{Testing}} and {{Failure Analysis}}}}.
  \bibinfo{publisher}{{ASM International}}, \bibinfo{pages}{65--72}.
\newblock
\urldef\tempurl%
\url{https://doi.org/10.31399/asm.cp.istfa2014p0065}
\showDOI{\tempurl}


\bibitem[\protect\citeauthoryear{O'Flynn and Chen}{O'Flynn and Chen}{2015}]%
        {side-channel-power-analysis-of-an-aes-256-bootloader}
\bibfield{author}{\bibinfo{person}{Colin O'Flynn} {and}
  \bibinfo{person}{Zhizhang~David Chen}.} \bibinfo{year}{2015}\natexlab{}.
\newblock \showarticletitle{Side channel power analysis of an AES-256
  bootloader}. In \bibinfo{booktitle}{\emph{2015 IEEE 28th Canadian Conference
  on Electrical and Computer Engineering (CCECE)}}. IEEE,
  \bibinfo{pages}{750--755}.
\newblock


\bibitem[\protect\citeauthoryear{Rahman, Tajik, Rahman, Tehranipoor, and
  Asadizanjani}{Rahman et~al\mbox{.}}{2020}]%
        {rahman_key_2020}
\bibfield{author}{\bibinfo{person}{Mir~Tanjidur Rahman},
  \bibinfo{person}{Shahin Tajik}, \bibinfo{person}{M.~Sazadur Rahman},
  \bibinfo{person}{Mark Tehranipoor}, {and} \bibinfo{person}{Navid
  Asadizanjani}.} \bibinfo{year}{2020}\natexlab{}.
\newblock \showarticletitle{The {{Key}} is {{Left}} under the {{Mat}}: {{On}}
  the {{Inappropriate Security Assumption}} of {{Logic Locking Schemes}}}. In
  \bibinfo{booktitle}{\emph{2020 {{IEEE International Symposium}} on {{Hardware
  Oriented Security}} and {{Trust}} ({{HOST}})}}.
\newblock
\urldef\tempurl%
\url{https://doi.org/10.1109/HOST45689.2020.9300258}
\showDOI{\tempurl}


\bibitem[\protect\citeauthoryear{Sa{\ss}, Mitev, and Sadeghi}{Sa{\ss}
  et~al\mbox{.}}{2023}]%
        {oops-i-glitched-it-again}
\bibfield{author}{\bibinfo{person}{Marvin Sa{\ss}}, \bibinfo{person}{Richard
  Mitev}, {and} \bibinfo{person}{Ahmad-Reza Sadeghi}.}
  \bibinfo{year}{2023}\natexlab{}.
\newblock \showarticletitle{Oops..! I Glitched It Again! How to Multi-Glitch
  the Glitching-Protections on ARM TrustZone-M}.
\newblock \bibinfo{journal}{\emph{arXiv preprint arXiv:2302.06932}}
  (\bibinfo{year}{2023}).
\newblock


\bibitem[\protect\citeauthoryear{Schellenberg, Finkeldey, Richter,
  Sch{\"a}pers, Gerhardt, Hofmann, and Paar}{Schellenberg
  et~al\mbox{.}}{2015}]%
  {On-the-complexity-reduction-of-laser-fault-injection-campaigns-using-OBIC-measurements}
\bibfield{author}{\bibinfo{person}{Falk Schellenberg}, \bibinfo{person}{Markus
  Finkeldey}, \bibinfo{person}{Bastian Richter}, \bibinfo{person}{Maximilian
  Sch{\"a}pers}, \bibinfo{person}{Nils Gerhardt}, \bibinfo{person}{Martin
  Hofmann}, {and} \bibinfo{person}{Christof Paar}.}
  \bibinfo{year}{2015}\natexlab{}.
\newblock \showarticletitle{On the complexity reduction of laser fault
  injection campaigns using OBIC measurements}. In
  \bibinfo{booktitle}{\emph{2015 Workshop on Fault Diagnosis and Tolerance in
  Cryptography (FDTC)}}. IEEE, \bibinfo{pages}{14--27}.
\newblock


\bibitem[\protect\citeauthoryear{Selmke, Strieder, Heyszl, Freud, and
  Damm}{Selmke et~al\mbox{.}}{2021}]%
        {breaking-black-box-crypto-devices-using-laser-fault-injection}
\bibfield{author}{\bibinfo{person}{Bodo Selmke}, \bibinfo{person}{Emanuele
  Strieder}, \bibinfo{person}{Johann Heyszl}, \bibinfo{person}{Sven Freud},
  {and} \bibinfo{person}{Tobias Damm}.} \bibinfo{year}{2021}\natexlab{}.
\newblock \showarticletitle{Breaking black box crypto-devices using laser fault
  injection}. In \bibinfo{booktitle}{\emph{International Symposium on
  Foundations and Practice of Security}}. Springer, \bibinfo{pages}{75--90}.
\newblock


\bibitem[\protect\citeauthoryear{Tajik, Lohrke, Seifert, and Boit}{Tajik
  et~al\mbox{.}}{2017}]%
        {tajik_power_2017}
\bibfield{author}{\bibinfo{person}{Shahin Tajik}, \bibinfo{person}{Heiko
  Lohrke}, \bibinfo{person}{Jean-Pierre Seifert}, {and}
  \bibinfo{person}{Christian Boit}.} \bibinfo{year}{2017}\natexlab{}.
\newblock \showarticletitle{On the {{Power}} of {{Optical Contactless
  Probing}}: {{Attacking Bitstream Encryption}} of {{FPGAs}}}. In
  \bibinfo{booktitle}{\emph{Proceedings of the 2017 {{ACM SIGSAC Conference}}
  on {{Computer}} and {{Communications Security}} ({{CCS}})}}.
  \bibinfo{publisher}{{ACM}}, \bibinfo{pages}{1661--1674}.
\newblock
\urldef\tempurl%
\url{https://doi.org/10.1145/3133956.3134039}
\showDOI{\tempurl}


\bibitem[\protect\citeauthoryear{Van~den Herrewegen, Oswald, Garcia, and
  Temeiza}{Van~den Herrewegen et~al\mbox{.}}{2021}]%
        {fill-your-boots}
\bibfield{author}{\bibinfo{person}{Jan Van~den Herrewegen},
  \bibinfo{person}{David Oswald}, \bibinfo{person}{Flavio~D Garcia}, {and}
  \bibinfo{person}{Qais Temeiza}.} \bibinfo{year}{2021}\natexlab{}.
\newblock \showarticletitle{Fill your boots: Enhanced embedded bootloader
  exploits via fault injection and binary analysis}.
\newblock \bibinfo{journal}{\emph{IACR Transactions on Cryptographic Hardware
  and Embedded Systems}} (\bibinfo{year}{2021}), \bibinfo{pages}{56--81}.
\newblock


\bibitem[\protect\citeauthoryear{Verbauwhede and Maes}{Verbauwhede and
  Maes}{2011}]%
        {physically-unclonable-functions}
\bibfield{author}{\bibinfo{person}{Ingrid Verbauwhede} {and}
  \bibinfo{person}{Roel Maes}.} \bibinfo{year}{2011}\natexlab{}.
\newblock \showarticletitle{Physically unclonable functions: manufacturing
  variability as an unclonable device identifier}. In
  \bibinfo{booktitle}{\emph{Proceedings of the 21st edition of the great lakes
  symposium on Great lakes symposium on VLSI}}. \bibinfo{pages}{455--460}.
\newblock


\end{thebibliography}
